\documentclass[final,3p,times]{elsarticle}

\usepackage{amssymb}

\usepackage{amsthm}

\begin{document}

\begin{frontmatter}

\title{ Quantum-gravitational running/reduction of space-time dimension \tnoteref{label1}\tnotetext[label1]{ This essay received Honorable Mention in the 2009 Awards for Essays in Gravitation by the Gravity Research Foundation.} }

\author{ Michael~Maziashvili }
\ead{mishamazia@hotmail.com}

\address{Andronikashvili
Institute of Physics, 6 Tamarashvili St., Tbilisi 0177, Georgia}

\address{Faculty of Physics and Mathematics, Chavchavadze
State University, \\ 32 Chavchavadze Ave., Tbilisi 0179, Georgia}

\begin{abstract}

Quantum-gravity renders the space-time dimension to depend on the size of region; it monotonically increases with the size of region and asymptotically approaches four for large distances. This effect was discovered in numerical simulations of lattice quantum gravity in the framework of causal dynamical triangulation [hep-th/0505113] as well as in renormalization group approach to quantum gravity [hep-th/0508202]. However, along these approaches the interpretation and the physical meaning of the effective change of dimension at shorter scales  is not clear. Without invoking particular models in this essay we show that, box-counting dimension in face of finite resolution of space-time (generally implied by quantum
gravity) shows a simple way how both the qualitative and the quantitative features of this effect can be understood. In this way we derive a simple analytic expression of space-time dimension running, which implies the modification of Newton's inverse square law in a perfect agreement with the modification coming from one-loop gravitational radiative corrections.

\end{abstract}

\begin{keyword}
Quantum gravity \sep Effective dimension of Space-time. 

\PACS 04.60.-m \sep 



\end{keyword}

\end{frontmatter}

\section*{\textsf{ Essay }}

A new door for a versatile study of the quantum gravity phenomenology is open
by a profound quantum gravity effect of space-time dimension running/reduction
discovered recently in two different approaches to quantum gravity
\cite{AJLLR}. Two important questions arise in this regard. 
First we have to understand what the physical meaning of this effect is, and second, we should find an expression of dimension 
running, which will have more or less universal form irrespective to the particular approaches to quantum gravity. Both of these questions are left open and call for attention. These are the questions we address in this essay. Before addressing the questions we should be clear about the concept of gravitational running of dimension. We start from four dimensional Minkowskian space-time in which the matter fields leave but gravity is switched off. When gravity enteres the game the efective dimension of space-time (what observer can measure) appears
to depend on the size of region, it becomes smaller than four
at small scales and monotonically increases with increasing the size of region
\cite{AJLLR}. In order to get good understanding of this statement let us recall what the dimension is \cite{Falconer}.

Let us consider a subset $\mathcal{F}$ of four dimensional Minkowskian space. To determine the dimension of $\mathcal{F}$ first we have to introduce Euclidean metric on it, that is, to consider it as a subset of Euclidean space
$\mathbb{R}^4$. It is straitforward to Euclideanize the initial metric that makes $\mathcal{F}$ a subset of $\mathbb{R}^4$. Then let $l^4$ be a smallest box containing this
set, $\mathcal{F} \subseteq l^4$. For estimating the dimension of
$\mathcal{F}$ we have to cover it by $\epsilon^4$ cells and
counting the minimal number of such cells, $N(\epsilon)$, we
determine the dimension, $d \equiv \dim(\mathcal{F})$ as a limit
$d = d(\epsilon \rightarrow 0)$, where $n^{d(\epsilon)} = N $ and
$n = l/\epsilon$ \cite{Falconer}. This
definition is referred to as a box-counting dimension and can be written
in a more familiar form as
\[ d \,=\, \lim\limits_{\epsilon \rightarrow 0} {\ln N(\epsilon) \over
\ln {l \over \epsilon}}~.\]
Certainly, in the case when $\mathcal{F} = l^4$, by taking the
limit $d(\epsilon \rightarrow 0)$ we get the dimension to be $4$.
From the fact that we are talking about the dimension of a set
embedded into the four dimensional space, $\mathcal{F} \subset
\mathbb{R}^4$, it automatically follows that its dimension can not
be greater than $4$, $d \le 4$. We see that the volume of a
fractal $\mathcal{F}$ uniformly filling the box $l^4$ is reduced in comparison with the
four dimensional value $l^4$
\begin{equation}\label{volumefluct} V(\mathcal{F}) \,=\, \lim\limits_{\epsilon \rightarrow 0} N(\epsilon)\epsilon^4
\,=\, \lim\limits_{\epsilon \rightarrow
0}n(\epsilon)^{d(\epsilon)}\epsilon^4 \,\leq \, l^4 ~.\end{equation}

So, in a generic mathematical treatment of dimension 
we measure a set in a way that ignores irregularities of size less
than $\epsilon$, and then we see how this measurement behaves as
$\epsilon \rightarrow 0$. A glance at a recent
physics literature shows the variety of natural objects that are
described as fractals $-$ cloud boundaries, topographical
surfaces, coastlines, turbulence in fluids, and so on. Such a description is motivated by the fact that in any real measurement we use a finite resolution $\epsilon$. In this case the dimension takes the form $(l/\epsilon)^{d(\epsilon)} = N(\epsilon) $, which after introducing $\varepsilon(\epsilon) = 4 - d(\epsilon)\,,~~\delta N =
n(\epsilon)^4 - N(\epsilon)$ can be written as \begin{equation}\label{opdim} \varepsilon(\epsilon) \,=\, - {\ln
\left(1 - {\delta N(\epsilon) \over n(\epsilon)^4 } \right)\over
\ln n(\epsilon) } \, \approx \, {1 \over \ln n(\epsilon)}
\,{\delta N(\epsilon) \over n(\epsilon)^4}~. \end{equation} However, it is widely held belief that none of these objects are actual fractals $-$ that their fractal features disappear if
they are viewed at sufficiently small scales. But this naive
expectation is impeded by quantum gravity. Quantum gravity strongly indicates the finite resolution of
space-time. A glance at a recent
quantum gravity literature obviously shows that the finite resolution of
space-time is common for all approaches to quantum gravity: space-time
uncertainty relations in string theory \cite{String1, String2};
noncommutative space-time approach \cite{noncommutative}; loop
quantum gravity \cite{Loop}; or space-time uncertainty relations
coming from a simple {\tt Gedankenexperiments} of space-time
measurement \cite{Gedankenexperiments}. Well known entropy bounds
emerging via the merging of quantum theory and general relativity
also imply finite space-time resolution \cite{entropybounds}. The
combination of quantum theory and general relativity in one or
another way manifests that the conventional notion of distance
breaks down the latest at the Planck scale $l_P \simeq
10^{-33}$\,cm \cite{minimumlength}.

So, in quantum gravity the resolution of space-time is set by the
Planck length $\epsilon = l_P$. The local fluctuations
$\sim l_P$ add up over the length scale $l$ to $\delta
l = (l_Pl)^{1/2}$ \cite{camelia, mazia}. Respectively, for the
region $l^4$ we get the volume fluctuation of the
order $\delta V = \delta l^4$, or in other words, in quantum gravity we expect
the Poison fluctuation of volume $l^4$ of the order $ \delta V =
(l^2/l_P^2)\, l_P^4 $ \cite{Sorkin}. In view of Eq.(\ref{volumefluct}) one
naturally finds that this volume fluctuation has to account for
the reduction of dimension.\footnote{ This suggestion has been
made in \cite{mazia}, though the rate of volume fluctuation was
overestimated in this paper. Let us also notice
that the necessity of operational definition of dimension because
of quantum mechanical uncertainties (not quantum\,-\,gravitational
!) was first stressed in \cite{SZ}.} Thus, using the Eq.(\ref{opdim}) one gets $n = l/l_P\,,~\delta N = l^2/l_P^2$\,,
\begin{equation}\label{rundim} \varepsilon (l) \,=\,{1 \over \ln {l\over l_P}}\,\, 
{\delta V \over l^4}  \,=\, {1 \over \ln {l\over l_P}}\,
\left({l_P \over l}\right)^2~.\end{equation}
This equation (which was obtained on general grounds) gives the running of dimension with
respect to the size of region $l$. Having found this equation, the question of its reliability naturally occurs. For this reason we consider the following particular example. 

In the low energy regime ($\ll m_P$) general relativity can be
successfully treated as an effective quantum field theory
\cite{Donoghue}. So that it is possible to unambiguously compute
quantum effects due to graviton loops, as long as the momentum of
the particles in the loops is cut off at some scale $\ll m_P$. The
results are independent of the structure of any ultraviolet
completion, and therefore constitute genuine low energy
predictions of any quantum theory of gravity. Following this way
of reasoning it has been possible to compute one-loop quantum
correction to the Newtonian potential \cite{Khriplovich}

\begin{equation}\label{oneloop}V(r) = -l_P^2{m_1\,m_2 \over r}\left[1 + \frac{41}{10\pi}  \left({l_P \over r}\right)^2 \right]
~.\end{equation} Let us compare this result with the modification of the Newton's law due to dimension running Eq.(\ref{rundim}).

Modification of the Newton's law due to dimension reduction can be
estimated without too much trouble by writing it in the Planck
units

\[ -l_Pm_1\,m_2 \,{ l_P \over r} \equiv  -l_Pm_1\,m_2 \,{ 1 \over \xi} ~. \] One easily finds

\[ { 1 \over \xi^{1 - \varepsilon}}  = {e^{\varepsilon\ln\xi} \over \xi} =  {1 \over \xi} \left(1 +  \varepsilon\ln\xi +
\ldots~\right)~.\] Substituting the Eq.(\ref{rundim})
\[ \varepsilon = {1 \over \xi^2\ln\xi} ~,\] we get

\[V(r) = -l_P^2{m_1\,m_2 \over r}\left[1 +  \left({l_P \over r}\right)^2 \right] ~,
\] which appears to be in perfect agreement with the Eq.(\ref{oneloop}).

\section*{ Acknowledgments }

Author acknowledges useful discussions with Zurab Silagadze. The work
was supported in part by the \emph{CRDF/GRDF} grant.

\end{document}